\documentclass[apjl]{emulateapj}
\usepackage{apjfonts}
\usepackage{multirow}
\usepackage{color}
\def\apjl{ApJ}%
\def\apjs{ApJS}%
\def\ao{Appl.~Opt.}%
\def\aap{A\&A}%

\defcitealias{AGS05}{AGS05}
\defcitealias{GS98}{GS98}
\definecolor{SN}{rgb}{0.,0.5,0.}
\definecolor{rftwo}{rgb}{0.5,0.,0.}

\slugcomment{Published in ApJL (2009) Vol. 691, L119-L122}

\shorttitle{The solar Ni and O abundances}
\shortauthors{Scott et al.}

\begin{document}

\newcommand{\SN}[1]{\textbf{\textcolor{SN}{#1}}}
\newcommand{\rftwo}[1]{\textbf{\textcolor{rftwo}{#1}}}

\title{On the solar nickel and oxygen abundances}

\author{Pat Scott,\altaffilmark{1,2,3} Martin Asplund,\altaffilmark{3,4} Nicolas Grevesse\altaffilmark{5,6} and A.~Jacques Sauval\altaffilmark{7}}
\altaffiltext{1}{Cosmology, Particle Astrophysics and String Theory, Department of Physics, Stockholm University, AlbaNova University Centre, SE-106 91 Stockholm, Sweden; pat@fysik.su.se} 
\altaffiltext{2}{Oskar Klein Centre for Cosmoparticle Physics}
\altaffiltext{3}{Most results were obtained whilst these authors were at the Australian National University Research School of Astronomy and Astrophysics, Mt. Stromlo Observatory, Cotter Rd., Weston Creek, ACT 2611, Australia.}
\altaffiltext{4}{Max Planck Institute for Astrophysics, Postfach 1317, D-85741 Garching b. M\"unchen, Germany; asplund@mpa-garching.mpg.de}
\altaffiltext{5}{Centre Spatial de Li{\`e}ge, Universit{\'e} de Li{\`e}ge, avenue Pr{\'e} Aily, B-4031 Angleur-Li{\`e}ge, Belgium} 
\altaffiltext{6}{Institut d'Astrophysique et de G{\'e}ophysique, Universit{\'e} de Li{\`e}ge, All{\'e}e du 6 ao{\^u}t, 17, B5C, B-4000 Li{\`e}ge, Belgium}
\altaffiltext{7}{Observatoire Royal de Belgique, avenue circulaire, 3, B-1180 Bruxelles, Belgium}

\begin{abstract}
Determinations of the solar oxygen content relying on the neutral forbidden transition at 630{\,nm} depend upon the nickel abundance, due to a Ni\,\textsc{i} blend.  Here we rederive the solar nickel abundance, using the same {\it ab initio} 3D hydrodynamic model of the solar photosphere employed in the recent revision of the abundances of C, N, O and other elements.  Using 17 weak, unblended lines of Ni\,\textsc{i} together with the most accurate atomic and observational data available we find $\log\epsilon_{\rm Ni}=6.17\pm0.02\,\mathrm{(statistical)}\pm0.05\,\mathrm{(systematic)}$, a downwards shift of 0.06--0.08\,dex relative to previous 1D-based abundances.  We investigate the implications of the new nickel abundance for studies of the solar oxygen abundance based on the [O\,\textsc{i}] 630{\,nm} line in the quiet Sun.  Furthermore, we demonstrate that the oxygen abundance implied by the recent sunspot spectropolarimetric study of Centeno \& Socas-Navarro needs to be revised downwards from  $\log\epsilon_{\rm O}=8.86\pm0.07$ to $8.71\pm0.10$.  This revision is based on the new nickel abundance, application of the best available $gf$-value for the 630{\,nm} forbidden oxygen line, and a more transparent treatment of CO formation.  Determinations of the solar oxygen content relying on forbidden lines now appear to converge around $\log\epsilon_{\rm O}=8.7$.
\end{abstract}


\begin{deluxetable*}{l@{\ }l@{\hspace{2.5mm}}l@{\ }l@{\hspace{2.5mm}}c@{\hspace{2.5mm}}c@{\hspace{2.5mm}}c@{\hspace{2.5mm}}c@{\hspace{2.5mm}}c@{\hspace{2.5mm}}c@{\hspace{2.5mm}}c@{\hspace{2.5mm}}c@{\hspace{2.5mm}}c@{\hspace{2.5mm}}c@{\hspace{2.5mm}}c@{\hspace{2.5mm}}c@{\hspace{2.5mm}}c@{\hspace{2.5mm}}c@{\hspace{2.5mm}}c@{\hspace{2.5mm}}c@{\hspace{2.5mm}}c}
\centering
\tablecaption{List of neutral nickel lines \label{lines}}
\tablewidth{0pt}
\tablehead{
\multicolumn{4}{c}{Atomic levels} & Isotope & Wavelength & Ex. Pot.& $\log gf$ & 
$gf$ & $\log\gamma_{\rm rad}$ & $\sigma$ & $\alpha$ & W$_\lambda$ & $\log\epsilon_{\rm Ni}$ & Weight \\
\multicolumn{2}{c}{Lower} & \multicolumn{2}{c}{Upper} & & (nm, air) & (eV) & (eff.) & ref. & & & & (pm) & (3D) &
}
\startdata
 3d$^8$($^3$F)4s4p($^3$P) & $^5$G$_4$  & 3d$^9$($^2$D)4d & $^3$G$_5$    & & 474.01658 &  3.480 & -1.730 & WL& 7.899 & 844 & 0.281 & 1.60  &  6.18 & 1\vspace{1mm}\\
 \multirow{2}{*}{3d$^9$($^2$D)4p} & \multirow{2}{*}{$^3$P$_1$}           & \multirow{2}{*}{3d$^9$($^2$D)4d} & \multirow{2}{*}{$^3$P$_0$}  & $^{58}$Ni & 481.19772 &                          & -1.592 &  &  &  &  &  & \\\vspace{1mm}
 \phantom{($^2$D)4p $^3$P$_1$}          &  &  &  & $^{60}$Ni & 481.19926 &  \multirow{-2}{*}{3.658} & -2.006 & \multirow{-2}{*}{J03} & \multirow{-2}{*}{8.285} & \multirow{-2}{*}{-} & \multirow{-2}{*}{-} & \multirow{-2}{*}{2.12} &  \multirow{-2}{*}{6.20} & \multirow{-2}{*}{1} \\\vspace{1mm}
 3d$^8$($^3$F)4s4p($^3$P) & $^5$G$_2$  & 3d$^8$4s($^4$F)5s & $^5$F$_3$  & & 481.45979 &  3.597 & -1.620 & WL& 8.053 & 743 & 0.236 & 1.58  &  6.17 & 1 \\\vspace{1mm}
 3d$^8$($^3$F)4s4p($^3$P) & $^5$G$_3$  & 3d$^8$4s($^4$F)5s & $^5$F$_4$  & & 487.47929 &  3.543 & -1.450 & WL& 8.039 &   - &     - & 2.35  &  6.15 & 1 \\\vspace{1mm}
 3d$^9$($^2$D)4p & $^3$D$_2$           & 3d$^8$4s($^4$F)5s & $^5$F$_2$  & & 488.67108 &  3.706 & -1.780 & WL& 8.211 &   - &     - & 0.90  &  6.13 & 1 \\\vspace{1mm}
 3d$^8$($^3$F)4s4p($^3$P) & $^5$G$_4$  & 3d$^8$4s($^4$F)5s & $^5$F$_5$  & & 490.09708 &  3.480 & -1.670 & WL& 8.062 & 693 & 0.238 & 1.79  &  6.17 & 1 \\\vspace{1mm}
 3d$^8$($^3$F)4s4p($^3$P) & $^5$F$_4$  & 3d$^9$($^2$D)4d & $^3$G$_4$    & & 497.61348 &  3.606 & -1.250 & WL& 7.962 & 843 & 0.282 & 2.86  &  6.13 & 2 \\\vspace{1mm}
 3d$^8$($^3$F)4s4p($^3$P) & $^5$F$_4$  & 3d$^8$4s($^4$F)5s & $^5$F$_5$  & & 515.79805 &  3.606 & -1.510 & WL& 8.093 & 691 & 0.236 & 1.86  &  6.13 & 3 \\\vspace{1mm}
 3d$^8$($^3$F)4s4p($^3$P) & $^3$G$_5$  & 3d$^8$4s($^4$F)5s & $^5$F$_4$  & & 550.40945 &  3.834 & -1.700 & WL& 8.063 & 713 & 0.240 & 0.97  &  6.18 & 1 \\\vspace{1mm}
 3d$^9$($^2$D)4p & $^1$F$_3$           & 3d$^9$($^2$D)4d & $^3$G$_4$    & & 551.00092 &  3.847 & -0.900 & WL& 8.215 &   - &     - & 3.76  &  6.17 & 2 \\\vspace{1mm}
 3d$^9$($^2$D)4p & $^1$F$_3$           & 3d$^8$4s($^4$F)5s & $^5$F$_4$  & & 553.71054 &  3.847 & -2.200 & WL& 8.280 & 695 & 0.216 & 0.31  &  6.16 & 3 \\
 \multirow{2}{*}{3d$^8$($^3$F)4s4p($^3$P)} & \multirow{2}{*}{$^3$G$_3$}  & \multirow{2}{*}{3d$^9$($^2$D)4d} & \multirow{2}{*}{$^3$G$_4$}  & $^{60}$Ni & 574.92795 &  & -2.526 &  &  &  &  &  &  & \\\vspace{1mm}
 \phantom{($^3$F)4s4p($^3$P) $^3$G$_3$} &  &  &  & $^{58}$Ni & 574.93039 &  \multirow{-2}{*}{3.941} & -2.112 &  \multirow{-2}{*}{WL} & \multirow{-2}{*}{7.944} & \multirow{-2}{*}{832} & \multirow{-2}{*}{0.284} & \multirow{-2}{*}{0.44} &  \multirow{-2}{*}{6.17} & \multirow{-2}{*}{2} \\
 \multirow{2}{*}{3d$^8$($^3$F)4s4p($^3$P)} & \multirow{2}{*}{$^3$F$_4$}  & \multirow{2}{*}{3d$^9$($^2$D)4d} & \multirow{2}{*}{$^3$G$_5$}  & $^{60}$Ni & 617.67980 &  & -0.816 &  &  &  &  &  &  & \\\vspace{1mm}
 \phantom{($^3$F)4s4p($^3$P) $^3$F$_4$} &  &  &  & $^{58}$Ni & 617.68200 &  \multirow{-2}{*}{4.088} & -0.402 &  \multirow{-2}{*}{WL} & \multirow{-2}{*}{8.162} & \multirow{-2}{*}{826} & \multirow{-2}{*}{0.284} & \multirow{-2}{*}{6.54} &  \multirow{-2}{*}{6.17} & \multirow{-2}{*}{2} \\\vspace{1mm}
 3d$^8$($^3$F)4s4p($^3$P) & $^3$F$_4$  & 3d$^8$4s($^4$F)5s & $^5$F$_4$  & & 620.46048 &  4.088 & -1.100 & WL& 8.244 & 719 & 0.247 & 2.11  &  6.19 & 3 \\
 \multirow{2}{*}{3d$^8$($^3$F)4s4p($^3$P)} & \multirow{2}{*}{$^3$F$_3$}  & \multirow{2}{*}{3d$^9$($^2$D)4d} & \multirow{2}{*}{$^3$G$_4$}  & $^{60}$Ni & 622.39710 &  & -1.466 &  &  &  &  &  &  & \\\vspace{1mm}
 \phantom{($^3$F)4s4p($^3$P) $^3$F$_3$} &  &  &  & $^{58}$Ni & 622.39914 &  \multirow{-2}{*}{4.105} & -1.052 &  \multirow{-2}{*}{WL} & \multirow{-2}{*}{8.322} & \multirow{-2}{*}{827} & \multirow{-2}{*}{0.283} & \multirow{-2}{*}{2.79} &  \multirow{-2}{*}{6.17} & \multirow{-2}{*}{3} \\
 \multirow{2}{*}{3d$^8$($^3$F)4s4p($^3$P)} & \multirow{2}{*}{$^3$D$_3$}  & \multirow{2}{*}{3d$^9$($^2$D)4d} & \multirow{2}{*}{$^3$G$_4$}  & $^{60}$Ni & 637.82328 &  & -1.386 &  &  &  &  &  &  & \\\vspace{1mm}
 \phantom{($^3$F)4s4p($^3$P) $^3$D$_3$} &  &  &  & $^{58}$Ni & 637.82580 &  \multirow{-2}{*}{4.154} & -0.972 &  \multirow{-2}{*}{WL} & \multirow{-2}{*}{8.317} & \multirow{-2}{*}{825} & \multirow{-2}{*}{0.283} & \multirow{-2}{*}{3.20} &  \multirow{-2}{*}{6.20} & \multirow{-2}{*}{3} \\
 3d$^8$($^3$F)4s4p($^3$P) & $^3$D$_3$  & 3d$^8$4s($^4$F)5s & $^5$F$_4$  & & 641.45884 &  4.154 & -1.180 & WL& 8.369 & 721 & 0.249 & 1.68  &  6.20 & 2\\
\enddata
\tablecomments{Wavelengths and excitation potentials are from \citet{Litzen93}.  Radiative damping is from VALD \protect\citep{VALD}.  Transition designations are from \citet{Wickliffe97} except in the case of 481.2\,nm, for which the designation is from VALD.  References for $gf$-values are WL: \protect\citet{Wickliffe97} and J03: \protect\citet{Johansson03}.  For lines with isotopic components $gf$-values are effective only, rescaled to reflect the (terrestrial) isotopic fractions of \protect\citet{IUPAC98}.  Collisional damping parameters $\alpha$ and $\sigma$ are courtesy of Paul Barklem (private communication, 1999; now in VALD, \citealt{Barklem00}).  Equivalent widths are from profile fits using the 3D model.}
\end{deluxetable*}

\section{Introduction}

The reference solar oxygen abundance has been revised over the past decade from $\log\epsilon_{\rm O}=8.93\pm0.04$ \citep{Anders89} via $8.83\pm0.06$ \citep[][\citetalias{GS98}]{GS98} to $8.66\pm0.05$ \citep[][\citetalias{AGS05}]{AGS05}.  This downward slide has been brought on by the tandem influences of three-dimensional photospheric models, treatment of departures from local thermodynamic equilibrium (LTE), identification of blends, improved atomic data and better observations \citep{APForbidO,AspIV}.  The new abundances of oxygen and other elements have solved many outstanding problems, but ruined agreement between helioseismological theory and observation \citep[see e.g.][]{Basu08}.  This has prompted a reanalysis of photospheric models, resulting in support for high \citep{Ayres06,Centeno08,Ayres08}, low \citep{Scott06,Socas07,Koesterke08,Melendez08} and intermediate \citep{Caffau08} solar oxygen abundances.

Many of these analyses rely upon the forbidden oxygen line at 630.0304\,{nm}, known to contain a significant blend from Ni\,\textsc{i} at \mbox{630.0341\,nm} \citep{APForbidO,Johansson03}.  The strength of this blend, and therefore the $\epsilon_{\rm O}$ indicated by [O\,\textsc{i}]\,\mbox{630\,nm}, depend critically upon the solar nickel abundance ($\epsilon_{\rm Ni}$).  This is no less true of the ingenious spectropolarimetric work of \citet{Centeno08} than of any other study based on [O\,\textsc{i}]\,\mbox{630\,nm}.  Here we accurately redetermine $\epsilon_{\rm Ni}$, and discuss the impact of the new value upon abundances from [O\,\textsc{i}]\,\mbox{630\,nm}.  We show that $\epsilon_{\rm Ni}$ is model-dependent, contradicting claims by \citeauthor{Centeno08} that their technique allows a nearly model-independent analysis of $\epsilon_{\rm O}$.

\section{Model atmospheres and observational data}
\label{models}

We used the same 3D LTE model atmosphere and line formation code as in earlier papers \citep[e.g.][]{AspII, AspIV}, described by \citet{AspI}.  We performed comparative calculations with three 1D models: HM \citep{HM}, \textsc{marcs} \citep{MARCS75, MARCS97} and 1DAV (a contraction of the 3D model into one dimension by averaging over surfaces of equal optical depth).  Each 1D model included a microturbulent velocity $\xi_{\rm t}=1\,\mathrm{km\,s}^{-1}$.  We averaged simulated intensity profiles over the temporal and spatial extent of the model atmosphere, and compared results with the Fourier Transform Spectrograph (FTS) disk-center atlas of \citet[][see also \citealt{Neckel99}]{Brault87}.  We removed the solar gravitational redshift of 633 m\,s$^{-1}$, and convolved simulated profiles with an instrumental sinc function of width $\Delta\sigma = \frac{c}{R}= 0.857$ km\,s$^{-1}$, reflecting the FTS resolving power $R=350\,000$ \citep{Neckel99}.  We obtained abundances with the 3D model from profile-fitting via a $\chi^2$-analysis, fitting local continua independently with nearby clear sections of the spectrum.  For 1D models we used the equivalent widths of 3D profile fits.

\section{Atomic data and line selection}
\label{atomicdata}

Our adopted Ni\,\textsc{i} lines and atomic data are given in Table~\ref{lines}.  The paucity of good lines and atomic data in the optical precludes any meaningful analysis of Ni\,\textsc{ii} in the Sun.  The most accurate Ni\,\textsc{i} oscillator strengths come from the laboratory FTS branching fractions (BFs) of \citet{Wickliffe97}, put on an absolute scale with the time-resolved laser-induced fluorescence (TRLIF) lifetimes of \citet{Bergeson93a}.  A small number of high-quality $gf$-values are also available from \citet{Johansson03}, based upon FTS BFs and a single TRLIF lifetime.  The uncertainties of the individual oscillator strengths we employ range from 0.02--0.07\,dex, but most are accurate to $\pm0.03$\,dex.

Nickel has five stable isotopes: $^{58}$Ni, $^{60}$Ni, $^{61}$Ni, $^{62}$Ni and $^{64}$Ni, present in the approximate ratio 74:28:1:4:1 in the Earth \citep{IUPAC98}.  Isotopic splitting of optical lines is small, and dominated by $^{58}$Ni and $^{60}$Ni.  \citet{Litzen93} have obtained accurate laboratory FTS wavelengths for the isotopic components of many Ni\,\textsc{i} lines, which we include in Table~\ref{lines} where applicable.  We model such lines with two components, distributing the total oscillator strength according to the terrestrial $^{58}$Ni:$^{60}$Ni ratio.  We are confident that the available data sufficiently describe the isotopic broadening of solar lines, as laboratory FTS recordings are far better resolved than lines in the Sun.  Isotopic structure has not been included in other determinations of the solar nickel abundance.

We gave weightings to lines from 1 to 3 according to the absence of blends in the solar spectrum and the clarity of the surrounding continuum.  Along with 16 unblended weak lines, we include the somewhat stronger \mbox{Ni\,\textsc{i} 617.7\,nm} (W$_\lambda=6.54$\,pm) owing to its clean line profile and its accurate atomic data.  This line should be less affected by the rigors of strong line formation than single-component lines of the same equivalent width, thanks to the desaturating effects of isotopic broadening.  To be conservative, we give this line a weighting of 2.  Our list has only two lines in common with \citet{Biemont80}, mainly due to the absence of accurate $gf$-values for the 10 other lines used by those authors.

We took wavelengths and excitation potentials from \citet{Litzen93}, and radiative damping from VALD \citep{VALD}.  For most lines, we used collisional broadening parameters calculated for individual lines by Paul Barklem (private communication, 1999; now in VALD, \citealt{Barklem00}).  For the remainder we interpolated in the tables of \citet{Anstee95} and \citet{Barklem97} where possible.  For \mbox{Ni\,\textsc{i} 488.7\,nm} we used the traditional \citet{Unsold} formula with a scaling factor of 2.0.  Transition designations are from \citet{Wickliffe97}, except for \mbox{Ni\,\textsc{i} 481.2\,nm}, where the designation is from VALD.  Apart from the slightly stronger 617.7\,nm line, all our lines are quite insensitive to the adopted collisional broadening.  The broadening treatment of \mbox{Ni\,\textsc{i} 488.7\,nm} thus has no impact on our abundance determination, nor does the ambiguity in the identification of the upper level of \mbox{Ni\,\textsc{i} 481.2\,nm} \citep{Johansson03}.

\begin{deluxetable*}{l c c c c c}
\centering
\tablecaption{Logarithmic solar nickel abundances (mean $\pm$ 1 standard deviation) \label{abuns}}
\tablewidth{0pt}
\tablehead{
& \colhead{3D} & \colhead{1DAV} & \colhead{HM} & \colhead{\textsc{marcs}} & \colhead{Meteoritic}
}
\startdata
$\log\epsilon_\mathrm{Ni}$ (Ni\,\textsc{i} lines) & $6.17\pm0.02$ & $6.17\pm0.02$ & $6.26\pm0.02$ & $6.16\pm0.02$ & $6.19\pm0.03$\\
\enddata
\end{deluxetable*}


\begin{figure}[tbp]
\centering
\vspace{1mm}
\includegraphics[width=0.84\linewidth]{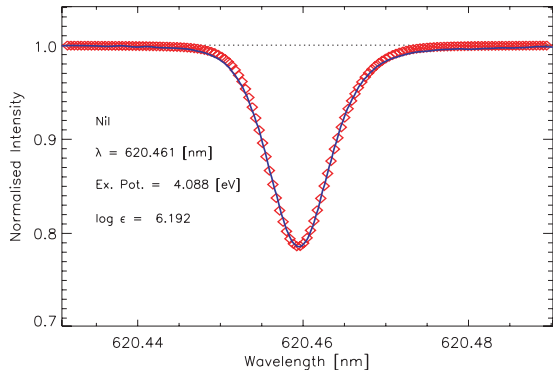}\vspace{2mm}
\includegraphics[width=0.84\linewidth]{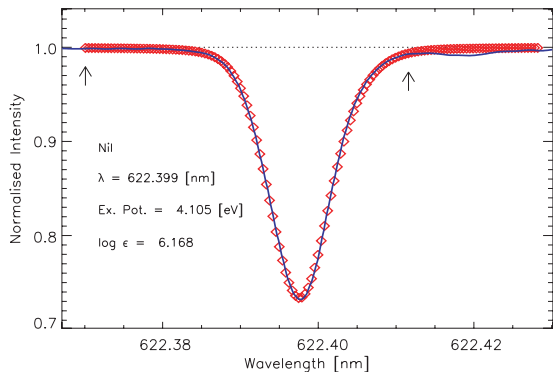}\vspace{2mm}
\includegraphics[width=0.84\linewidth]{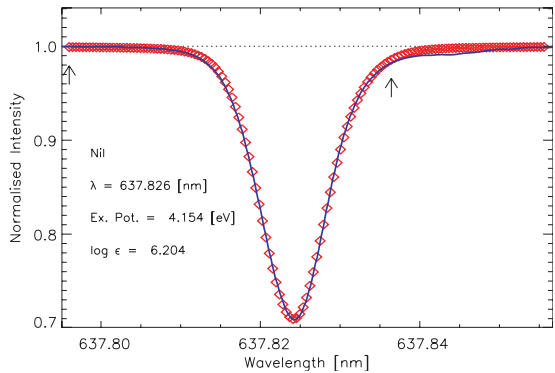}\vspace{2mm}
\caption[Example Ni\,\textsc{i} line profiles]{Example spatially- and temporally-averaged, disk-center synthesized Ni\,\textsc{i} line profiles (diamonds), compared with observed FTS profiles (solid lines).  We removed the solar gravitational redshift from the FTS spectrum, convolved synthesized profiles with instrumental sinc functions and fitted them in abundance and line shift.  Arrows indicate the windows used for profile fits; in the uppermost panel, we used the entire synthesized region.\label{profiles}}
\end{figure}

\begin{figure}[tbp]
\includegraphics[width=0.97\linewidth,trim = 0 0 0 15, clip=true]{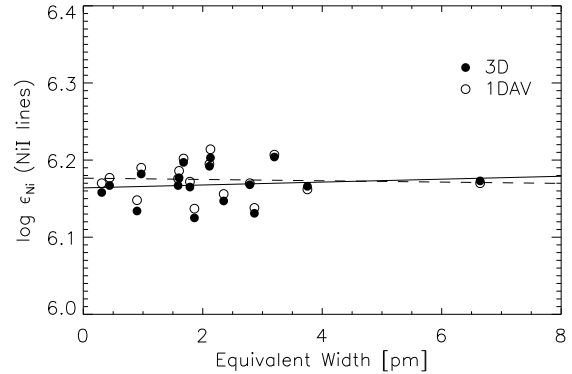}
\includegraphics[width=0.97\linewidth]{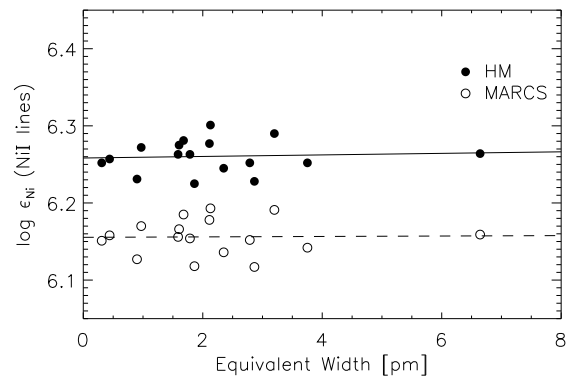}
\caption[Ni\,\textsc{i}-indicated Ni abundance plots]{Solar nickel abundances indicated by Ni\,\textsc{i} lines, as computed with different model atmospheres.  No substantial trends with line strength are visible with any of the four models.  The similarity between 3D and 1DAV abundances suggests that the mean temperature structure of the 3D model is predominantly responsible for the difference between HM and 3D results, rather than the presence of temperature inhomogeneities.\label{plots}}
\end{figure}

\section{Nickel results}
\label{results}

The mean nickel abundances we found using different model atmospheres are given in Table~\ref{abuns}.  Examples of profile fits to Ni\,\textsc{i} lines with the 3D model are given in Fig.~\ref{profiles}, exhibiting similarly impressive agreement with observation as seen with other species \citep[e.g.][]{AspI,AspIV}.  None of the models show abundance trends with equivalent width (Fig.~\ref{plots}), excitation potential nor wavelength, and the scatter is universally low, boding well for the internal consistency of all models.  Very little difference exists between 3D and 1DAV abundances, implying that the mean temperature structure rather than atmospheric inhomogeneities is the main reason for the difference between the 3D and HM results.  The 3D $\epsilon_{\rm Ni}$ is in excellent agreement with the meteoritic value \citepalias{AGS05}, whereas the HM value is not.

We adopt the 3D Ni\,\textsc{i} result as the best estimate of the solar abundance:
\begin{displaymath}
\epsilon_{\rm Ni}=6.17\pm0.05.
\end{displaymath}
The total error ($\pm$0.05\,dex) is the sum in quadrature of the line-to-line scatter ($\pm$0.02\,dex) and potential systematics arising from the model atmosphere ($\pm$0.05\,dex).  \citetalias{AGS05} gave $\epsilon_\mathrm{Ni}=6.23\pm0.04$, from \citet{Reddy03} using an \textsc{atlas9} model (Kurucz, \url{http://kurucz.harvard.edu/grids.html}).  Previous reviews \citepalias[e.g.][]{GS98} adopted $6.25\pm0.09$, by \citet{Biemont80} using the HM model.  Our value is 0.06--0.08\,dex lower than earlier ones, and 0.09\,dex less than our own HM-based estimate.  There is presently no evidence for non-LTE effects on our chosen lines in the Sun \citep{AspARAA}, but without a dedicated study we cannot rule them out.  After adjusting for $gf$-values and equivalent widths, we find abundances $0.06$ and $0.07$\,dex higher with the HM model than \citeauthor{Biemont80}\ for the two lines in common.  We have not been able to trace the exact cause of these disparities, but tentatively attribute them to differences in radiative transfer codes, continuum opacities and implementations of the HM model.

\section{Implications for the solar oxygen abundance}

The revised solar nickel abundance presented here has a direct impact upon any derivation of the oxygen abundance using the [O\,\textsc{i}]\,\mbox{630\,nm} line, as this line is blended with one from Ni\,\textsc{i} \citep{APForbidO}.

\citet{Centeno08} used the Stokes~$V$ profile of [O\,\textsc{i}]\,\mbox{630\,nm} to find an atomic ratio $\epsilon_{\rm O,atomic} / \epsilon_{\rm Ni} = 210\pm24$ in a sunspot.  They adopted an outdated $gf$-value for the [O\,\textsc{i}]\,\mbox{630\,nm} line \citep[cf.][]{Storey00}, causing an overestimation of the ratio by 15\% ($+0.06$\,dex).  They assumed $\log\epsilon_{\rm Ni}=6.23$ to find $\epsilon_{\rm O,atomic}$ and converted this to a bulk $\epsilon_{\rm O}$ by calculating that 51\% of oxygen resides in molecules.  This is a reasonable assumption; in sunspots the only significant oxygen-bearing molecule is CO, which (roughly) forms as many molecules as there are carbon atoms available, due to the low temperatures.  This number thus mirrors the assumed C/O ratio at the start of the calculation.  That ratio only depends weakly on the choice of 3D or HM model, as seen in the shift from $0.49\pm0.11$ to $0.54\pm0.10$ between \citetalias{GS98} and \citetalias{AGS05}.  A more straightforward way of estimating the contribution from CO would be to say that the maximum $\epsilon_{\rm CO}$ is given by the adopted carbon abundance: $\epsilon_{\rm O}\approx\epsilon_{\rm O,atomic}+\epsilon_{\rm C}$.

\citeauthor{Centeno08} claimed a nearly model-independent analysis because neither their CO correction nor nickel-to-atomic-oxygen ratio relied on an atmospheric model, and they believed $\epsilon_{\rm Ni}$ to be well-established.  The first statement is approximately true in the current debate, and the second is true of photospheric (but not sunspot) models.  Here we have shown that $\epsilon_{\rm Ni}$ is a model-dependent quantity, however.  The determination of $\epsilon_\mathrm{O}$ via \citeauthor{Centeno08}'s method is thus manifestly model-dependent, so there is no longer any reason to prefer placing a prior on the C/O ratio than on $\epsilon_\mathrm{C}$ directly.  Using our new nickel abundance, correcting the [O\,\textsc{i}]\,\mbox{630\,nm} $gf$, adopting the $\epsilon_{\rm C}$ of \citetalias{AGS05} and fully propagating all errors, we find an oxygen abundance of $\log\epsilon_{\rm O}=8.71\pm0.10$ instead of their $8.86\pm0.07$.  Had we adopted the traditional sunspot model of \citet{Maltby86} instead of the one inferred from spectrum inversion by \citeauthor{Centeno08}, we would have found $8.67\pm0.10$.  Retaining \citeauthor{Centeno08}'s prior on the C/O ratio (with the error thereupon given by \citetalias{AGS05}), one would obtain $8.74\pm0.10$ with their sunspot model and $8.66\pm0.10$ with the \citeauthor{Maltby86}\ model. Clearly their method is not as model-insensitive as \citeauthor{Centeno08} argued.

Our new Ni abundance also modifies analyses of [O\,\textsc{i}]\,\mbox{630\,nm} in the quiet solar spectrum.  \citet{APForbidO}, \citet{AspIV} and \citet{Ayres08} all allowed the Ni contribution to vary freely in their 3D profile-fitting of the 630\,nm feature, whilst \citet{Caffau08} fixed it with the $\epsilon_{\rm Ni}$ of \citetalias{GS98}.  With the Ni abundances from Table \ref{abuns} and the laboratory $gf$-value of the Ni\,\textsc{i} blend \citep{Johansson03}, we can now accurately predict the Ni contribution to [O\,\textsc{i}]\,\mbox{630\,nm}.  Independent of the adopted 1D or 3D model atmosphere, it is 0.17\,pm in disk-center intensity, and 0.19\,pm in flux.  In terms of oxygen abundance, this implies a decrease by 0.04\,dex to $\log\epsilon_{\rm O}\approx8.65$ for the analysis of \citet{AspIV}, further improving the excellent agreement between different indicators.  The derived abundance of \citet{Ayres08} would decrease to about 8.77 while that of \citet{Caffau08} would increase to approximately 8.72.  Because we now know the strength of the Ni blend, it is surprising that these two studies yield different results for the remaining contribution from oxygen, as they both rely on the same 3D CO$^5$BOLD model. Since \citeauthor{Caffau08} employed several 3D snapshots whereas \citeauthor{Ayres08} used only one, we tentatively consider the former more reliable.  No Ni abundance has yet been estimated with the CO$^5$BOLD model, but regardless of its value our conclusions about the strength of the Ni\,\textsc{i} 630\,nm blend, and thus its impact on oxygen abundances found by different authors, would remain unchanged.  The difference of approximately 0.07\,dex in the revised \citet{AspIV} and \citet{Caffau08} abundances from [O\,\textsc{i}]\,\mbox{630\,nm} probably reflects the different mean temperature stratifications of the two 3D models.

Given our reappraisal of the oxygen abundances of \cite{Centeno08}, \citet{AspIV} and \citet{Caffau08}, together with the recent study of \citet{Melendez08} using the [O\,\textsc{i}] 557.7\,nm line, it now seems that results from forbidden oxygen lines are beginning to converge around $\log\epsilon_\mathrm{O}=8.7$.  Whilst this agreement might come as a relief to some, it only serves to sharpen the current discrepancy between spectroscopy and helioseismology.

\acknowledgments
We would like to take this opportunity to commemorate the work and life of Sveneric Johansson, his contribution to atomic spectroscopy in general and to nickel and [O\,\textsc{i}]\,\mbox{630\,nm} in particular.  PS thanks IAU Commission 46, the ANU and the Australian Research Council for financial support.

\bibliography{AbuGen,FePeakGeneral,CandO,Ni,CObiblio}

\begin{thebibliography}{36}
\expandafter\ifx\csname natexlab\endcsname\relax\def\natexlab#1{#1}\fi

\bibitem[{{Allende Prieto} {et~al.}(2001){Allende Prieto}, {Lambert}, \&
  {Asplund}}]{APForbidO}
{Allende Prieto}, C., {Lambert}, D.~L., \& {Asplund}, M. 2001, \apjl, 556, L63

\bibitem[{{Anders} \& {Grevesse}(1989)}]{Anders89}
{Anders}, E., \& {Grevesse}, N. 1989, \gca, 53, 197

\bibitem[{{Anstee} \& {O'Mara}(1995)}]{Anstee95}
{Anstee}, S.~D., \& {O'Mara}, B.~J. 1995, \mnras, 276, 859

\bibitem[{{Asplund}(2005)}]{AspARAA}
{Asplund}, M. 2005, \araa, 43, 481

\bibitem[{{Asplund} {et~al.}(2005){Asplund}, {Grevesse}, \& {Sauval}}]{AGS05}
{Asplund}, M., {Grevesse}, N., \& {Sauval}, A.~J. 2005, in ASP Conf. Ser. 336,
  ed. T.~G. {Barnes III} \& F.~N. {Bash} (Astron. Soc. Pac., San Francisco),
  25, \citepalias{AGS05}

\bibitem[{{Asplund} {et~al.}(2004){Asplund}, {Grevesse}, {Sauval}, {Allende
  Prieto}, \& {Kiselman}}]{AspIV}
{Asplund}, M., {Grevesse}, N., {Sauval}, A.~J., {Allende Prieto}, C., \&
  {Kiselman}, D. 2004, \aap, 417, 751

\bibitem[{{Asplund} {et~al.}(1997){Asplund}, {Gustafsson}, {Kiselman}, \&
  {Eriksson}}]{MARCS97}
{Asplund}, M., {Gustafsson}, B., {Kiselman}, D., \& {Eriksson}, K. 1997, \aap,
  318, 521

\bibitem[{{Asplund} {et~al.}(2000{\natexlab{a}}){Asplund}, {Nordlund},
  {Trampedach}, {Allende Prieto}, \& {Stein}}]{AspI}
{Asplund}, M., {Nordlund}, {\AA}., {Trampedach}, R., {Allende Prieto}, C., \&
  {Stein}, R.~F. 2000{\natexlab{a}}, \aap, 359, 729

\bibitem[{{Asplund} {et~al.}(2000{\natexlab{b}}){Asplund}, {Nordlund},
  {Trampedach}, \& {Stein}}]{AspII}
{Asplund}, M., {Nordlund}, {\AA}., {Trampedach}, R., \& {Stein}, R.~F.
  2000{\natexlab{b}}, \aap, 359, 743

\bibitem[{{Ayres}(2008)}]{Ayres08}
{Ayres}, T.~R. 2008, \apj, 686, 731

\bibitem[{{Ayres} {et~al.}(2006){Ayres}, {Plymate}, \& {Keller}}]{Ayres06}
{Ayres}, T.~R., {Plymate}, C., \& {Keller}, C.~U. 2006, \apjs, 165, 618

\bibitem[{{Barklem} \& {O'Mara}(1997)}]{Barklem97}
{Barklem}, P.~S., \& {O'Mara}, B.~J. 1997, \mnras, 290, 102

\bibitem[{{Barklem} {et~al.}(2000){Barklem}, {Piskunov}, \&
  {O'Mara}}]{Barklem00}
{Barklem}, P.~S., {Piskunov}, N., \& {O'Mara}, B.~J. 2000, \aaps, 142, 467

\bibitem[{{Basu} \& {Antia}(2008)}]{Basu08}
{Basu}, S., \& {Antia}, H.~M. 2008, \physrep, 457, 217

\bibitem[{{Bergeson} \& {Lawler}(1993)}]{Bergeson93a}
{Bergeson}, S.~D., \& {Lawler}, J.~E. 1993, \josab, 10, 794

\bibitem[{{Bi{\'e}mont} {et~al.}(1980){Bi{\'e}mont}, {Grevesse}, {Huber}, \&
  {Sandeman}}]{Biemont80}
{Bi{\'e}mont}, E., {Grevesse}, N., {Huber}, M.~C.~E., \& {Sandeman}, R.~J.
  1980, \aap, 87, 242

\bibitem[{{Brault} \& {Neckel}(1987)}]{Brault87}
{Brault}, J., \& {Neckel}, H. 1987, {Spectral atlas of solar absolute
  disk-averaged and disk-centre intensity from 3290 to 12510\AA}
  ({ftp://ftp.hs.uni-hamburg.de/pub/outgoing/FTS-Atlas})

\bibitem[{{Caffau} {et~al.}(2008){Caffau}, {Ludwig}, {Steffen}, {Ayres},
  {Bonifacio}, {Cayrel}, {Freytag}, \& {Plez}}]{Caffau08}
{Caffau}, E., {Ludwig}, H.-G., {Steffen}, M., {Ayres}, T.~R., {Bonifacio}, P.,
  {Cayrel}, R., {Freytag}, B., \& {Plez}, B. 2008, \aap, 488, 1031

\bibitem[{{Centeno} \& {Socas-Navarro}(2008)}]{Centeno08}
{Centeno}, R., \& {Socas-Navarro}, H. 2008, \apjl, 682, L61

\bibitem[{{Grevesse} \& {Sauval}(1998)}]{GS98}
{Grevesse}, N., \& {Sauval}, A.~J. 1998, \ssr, 85, 161, \citepalias{GS98}

\bibitem[{{Gustafsson} {et~al.}(1975){Gustafsson}, {Bell}, {Eriksson}, \&
  {Nordlund}}]{MARCS75}
{Gustafsson}, B., {Bell}, R.~A., {Eriksson}, K., \& {Nordlund}, {\AA}. 1975,
  \aap, 42, 407

\bibitem[{{Holweger} \& {M\"uller}(1974)}]{HM}
{Holweger}, H., \& {M\"uller}, E.~A. 1974, \solphys, 39, 19

\bibitem[{{Johansson} {et~al.}(2003){Johansson}, {Litz{\'e}n}, {Lundberg}, \&
  {Zhang}}]{Johansson03}
{Johansson}, S., {Litz{\'e}n}, U., {Lundberg}, H., \& {Zhang}, Z. 2003, \apjl,
  584, L107

\bibitem[{{Koesterke} {et~al.}(2008){Koesterke}, {Allende Prieto}, \&
  {Lambert}}]{Koesterke08}
{Koesterke}, L., {Allende Prieto}, C., \& {Lambert}, D.~L. 2008, \apj, 680, 764

\bibitem[{{Kupka} {et~al.}(1999){Kupka}, {Piskunov}, {Ryabchikova}, {Stempels},
  \& {Weiss}}]{VALD}
{Kupka}, F., {Piskunov}, N., {Ryabchikova}, T.~A., {Stempels}, H.~C., \&
  {Weiss}, W.~W. 1999, \aaps, 138, 119

\bibitem[{{Litzen} {et~al.}(1993){Litzen}, {Brault}, \& {Thorne}}]{Litzen93}
{Litzen}, U., {Brault}, J.~W., \& {Thorne}, A.~P. 1993, \physscr, 47, 628

\bibitem[{{Maltby} {et~al.}(1986){Maltby}, {Avrett}, {Carlsson},
  {Kjeldseth-Moe}, {Kurucz}, \& {Loeser}}]{Maltby86}
{Maltby}, P., {Avrett}, E.~H., {Carlsson}, M., {Kjeldseth-Moe}, O., {Kurucz},
  R.~L., \& {Loeser}, R. 1986, \apj, 306, 284

\bibitem[{{Mel{\'e}ndez} \& {Asplund}(2008)}]{Melendez08}
{Mel{\'e}ndez}, J., \& {Asplund}, M. 2008, \aap, 490, 817

\bibitem[{{Neckel}(1999)}]{Neckel99}
{Neckel}, H. 1999, \solphys, 184, 421

\bibitem[{{Reddy} {et~al.}(2003){Reddy}, {Tomkin}, {Lambert}, \& {Allende
  Prieto}}]{Reddy03}
{Reddy}, B.~E., {Tomkin}, J., {Lambert}, D.~L., \& {Allende Prieto}, C. 2003,
  \mnras, 340, 304

\bibitem[{Rosman \& Taylor(1998)}]{IUPAC98}
Rosman, K. J.~R., \& Taylor, P. D.~P. 1998, Pure \& Appl. Chem., 70, 217

\bibitem[{{Scott} {et~al.}(2006){Scott}, {Asplund}, {Grevesse}, \&
  {Sauval}}]{Scott06}
{Scott}, P.~C., {Asplund}, M., {Grevesse}, N., \& {Sauval}, A.~J. 2006, \aap,
  456, 675

\bibitem[{{Socas-Navarro} \& {Norton}(2007)}]{Socas07}
{Socas-Navarro}, H., \& {Norton}, A.~A. 2007, \apjl, 660, L153

\bibitem[{{Storey} \& {Zeippen}(2000)}]{Storey00}
{Storey}, P.~J., \& {Zeippen}, C.~J. 2000, \mnras, 312, 813

\bibitem[{{Uns\"old}(1955)}]{Unsold}
{Uns\"old}, A. 1955, {Physik der Sternatmospharen, MIT besonderer
  Berucksichtigung der Sonne.}, 2nd edn. (Springer, Berlin)

\bibitem[{{Wickliffe} \& {Lawler}(1997)}]{Wickliffe97}
{Wickliffe}, M.~E., \& {Lawler}, J.~E. 1997, \apjs, 110, 163

\end{thebibliography}

\end{document}